\documentclass[journal]{IEEEtran}

\ifCLASSINFOpdf
\usepackage[pdftex]{graphicx}

\else

\fi

\usepackage{verbatim}
\usepackage[fleqn]{amsmath}

\usepackage[square, comma, sort&compress,numbers]{natbib}
\usepackage[table]{xcolor}  
\usepackage{tabu}
\usepackage{color}
\usepackage{flushend}
\usepackage{algorithm}
\usepackage{amssymb} 
\usepackage{graphicx}
\usepackage{url}
\graphicspath{{figures/}}
\DeclareGraphicsExtensions{.pdf,.jpeg,.png}

\usepackage{times}

\usepackage[center]{caption}

\usepackage{url}

\makeatletter
\def\url@leostyle{%
  \@ifundefined{selectfont}{\def\UrlFont{\sf}}{\def\UrlFont{\small\ttfamily}}}
\makeatother
\begin{document}
%

\title{Dynamic Co-Simulation Methods for \\ Combined Transmission-Distribution System and \\ Integration Time Step Impact on Convergence} 

\author{Ramakrishnan Venkatraman, \textit{Student Member, IEEE},~~~ Siddhartha~Kumar~Khaitan, \textit{Senior Member, IEEE} and Venkataramana Ajjarapu, \textit{Fellow, IEEE}

\thanks{The authors are with Dept. of ECE at Iowa State University, Ames, IA, USA. (e-mail:rvenkat@iastate.edu, skhaitan@iastate.edu; vajjarap@iastate.edu). This work was supported by the Dept. of Energy and the Electric Power Research Center at Iowa State University.}
}

\maketitle


\begin{abstract}
	Combined Transmission and Distribution Systems (CoTDS) simulation for power systems requires development of algorithms and software that are numerically stable and at the same time accurately simulate dynamic events that can occur in practical systems. The dynamic behavior of transmission and distribution systems are vastly different, especially with the increased deployment of distribution generation. The time scales of simulation can be orders of magnitude apart making the combined simulation extremely challenging. This has led to increased research in applying co-simulation techniques for integrated simulation of the two systems. In this paper, a rigorous mathematical analysis on convergence of numerical methods in co-simulation is presented. Two methods for co-simulation of CoTDS are proposed using parallel and series computation of the transmission system and distribution systems. Both these co-simulation methods are validated against total system simulation in a single time-domain simulation environment. The series computation co-simulation method is shown to have better numerical stability at larger integration time steps. The series computation co-simulation method is additionally validated against commercial EMTP software and the results show remarkable correspondence.	
\end{abstract}


 \newcommand\T{\rule{0pt}{2.6ex}}       
 \newcommand\B{\rule[-1.2ex]{0pt}{0pt}} 
 

\section{Introduction}
\label{Intro}
The modern distribution systems are becoming more active 
with the increased deployment of Distribution Generation (DG), especially with power-electronic inverters and smart grid control technologies which add a new dimension to system dynamics. From the transmission system perspective, the NERC standard TPL-001-4 \cite{NERC1}, Section R2.4.1 states that: ``System peak load levels shall include a load model which represents the expected dynamic behavior of loads that could impact the study area, considering the behavior of induction motor loads."  These factors have necessitated a new interest amongst researchers in modeling system dynamics for an integrated simulation of transmission and distribution systems and to develop algorithms that are numerically stable and at the same time accurately simulate dynamic events that can occur in practical systems.

Combined simulation also referred to as co-simulation can broadly be classified into three categories. First \cite{Sun2015master, Balasubramanium2017combined}, is the capability to solve the power flow of the two systems for obtaining a steady state power flow solution which is particularly suited for optimal power flow, planning algorithms, etc. Here, generally there are no dynamics involved. The second category  \cite{Kalsi2013integrated, Hale2015integrated, Palensky2017applied}, is to integrate the controls and communications into the steady state power flows for energy management and market clearance. In this category, the dynamics of the system are related to hourly or daily load and generation profiles. The third category, which is the focus area of this paper, deals with the detailed transient behavior and the interaction of the two systems.

It is quite a challenging task to simulate the transient behavior with both the transmission and distribution systems. Conventional single simulators for transmission and distribution systems have been developed and optimized over several years, and applying them to combined simulation often compromises the numerical behavior \cite{Palensky2017applied}. Currently, there are software like PSS\textregistered E, PSLF, and PSAT for the transmission system analysis and tools like OpenDSS and Gridlab-D for the distribution system analysis. But there are hardly any efficient commercial software for combined study of transmission and distribution systems with the exception of time-consuming Electro-Magnetic Transient Program (EMTP) simulators like EMTP-RV, PSCAD and MATLAB Simscape PowerSystems. Software tools such as DigSILENT, PSS\textregistered SINCAL is capable of simulations in different time scales but uses EMT simulation function for detailed dynamic simulation. 

Recent developments in co-simulation include GridMat \cite{Abdullah20xxGridMat}, Bus.py \cite{Hansen2015Buspy}, FNCS \cite{FNCS} which mainly cater to the first and second categories. 
One approach towards transient co-simulation is to use a combination of Transient-Stability type as the main simulator and embedding an EMTP type simulator by an inner calculation loop. In literature, co-simulation of two network systems for transient analysis using this approach are presented in \cite{Wenzhuo2011electromechanical} by integrating electromechanical and EMT simulation of transmission systems. The concept is to perform detailed study on a small part of a large system by dividing the whole system into external phasor domain network and detailed internal networks which interface through Th{\'e}venin and Norton equivalents at the boundary. This work is extended in  \cite{Huang2015application, Huang2016OpenHybridSim} where an EMT-Transient Stability hybrid simulation architecture is proposed.
The method is effective but still requires computationally intensive EMTP for the detailed internal network. A similar approach  extending to a frequency dependent network equivalent is presented in \cite{Zhang2015implicitly}. 

In \cite{Aristidou2013dynamic}, dynamic simulation of combined transmission and distribution systems is introduced to address the computational burden of representing all distribution networks in detail. A domain decomposition approach based on level of participation of distribution networks in system dynamics is adopted to distinguish between selecting a simple or detailed model. The networks are however, still solved using the complete set of differential algebraic equations in phasor domain.

In \cite{Oh2016cosimulation}, a co-simulation framework by two
independent EMT simulations with a time-delay compensation algorithm is proposed to improve the co-simulation accuracy, but is not suitable for large distribution networks.
In \cite{Jain2015three}, a novel three-phase dynamic analyzer algorithm is presented  that enables the study of  electromechanical transients in unbalanced networks without using EMTP programs. The idea behind this approach is to accurately simulate electromechanical transients using 3-phase approach. However, the method actually solves the system's differential equations in $dqo$ reference frame for instantaneous values and recovers the $abc$ values to solve the network algebraic equations and so the solution, although maintaining higher accuracy will inherently exhibit higher simulation times.

In \cite{Huang2017OpenSource}, an open-source co-simulation framework 
(FNCS), is introduced for managing the interaction and synchronisation of the
transmission and distribution simulators. The concept of dynamic co-simulation presented in this paper is highly relevant to the ongoing research in this area. However, there is no detailed analysis on the convergence aspect of dynamic co-simulation. The authors mention that they use a small simulation time step to avoid numerical errors and non-convergence problems.  

The principal issue in co-simulation is that the dynamic components in bulk transmission and in distribution systems can have different time constants. To accurately capture the dynamics, the integration time step chosen for the whole system must be according to the smallest time constant which makes the whole simulation very slow. In addition when the distribution system differential algebraic  equations are solved as an entire sub-system it can make the numerical solution tedious and cumbersome. To address the needs of dynamic CoTDS modeling,  we  briefly introduced a new co-simulation method in a conference paper \cite{venkatraman2017combined}. But the aspects of stability and convergence of the numerical method was not described. The impact of the integration time step in co-simulation is also not studied in detail in any of the existing literature.

The main contributions in this paper are 1) A detailed mathematical formulation of two prominent algorithms for co-simulation methods, i.e., series and parallel computation methods are presented for solving two blocks of differential-algebraic equations (DAEs) algebraically coupled to each other. The impact of integration time-step on stability and convergence is discussed. 2) Based on these co-simulation methods, two CoTDS dynamic simulation methods are proposed using existing single-phase transmission system dynamic solver and the three-phase distribution system power flow solver with an added interface to handle the distribution node-level dynamic components. 3) The proposed methods for CoTDS dynamic simulation are conceptually validated against a total system solution using a single phasor-domain simulation and also with EMTP simulation.

The rest of the paper is organized as follows: In Section II,  the mathematical background for convergence in co-simulation of coupled systems is presented. In Section III, two co-simulation methods for CoTDS dynamic simulation is proposed and these methods are validated in Section IV.
The paper is concluded in Section V.

\section{Mathematical Background for Convergence in Co-Simulation of Coupled Systems}
\label{sec:Math}
Simulation of a system that consists of well described sub-systems by using appropriate solvers for each sub-system is desired. To couple two or more sub-system solvers in time domain, co-simulation methods are used. In co-simulation the sub-systems are solved separately and the immediate mutual influence of subsystems is replaced by exchanging data at fixed time points \cite{Thilo2017Onmeeting}. In this section, the co-simulation concept is discussed for series and parallel computation of sub-systems and a convergence analysis of these methods is presented.

\subsection{Preliminaries}
\label{sec:Math_Pre}
Let us first consider a standard ordinary differential equation (\textit{ode}) in state variable $x$ given by 
\begin{equation}
\dot{x} = f(t,x)
\end{equation}
Convergence of numerical integration methods of such an \textit{ode} are generally analyzed using the following definitions \cite{Holmes2006IntroNumerical}.

\textit{Definition 1- Consistency}: A numerical method is called \textit{consistent} if the local truncation error, $\tau_i$ at the $i$th time step given by equation (\ref{eq:consistency}), approaches 0  as the time step, $H$ $\rightarrow$ 0.
\begin{equation}
\begin{aligned}
\tau(t_i,x_i,H) &= \frac{x(t_{i+1}) - x(t_i)}{H}  - \phi(t_i,x_i,H)
\end{aligned}
\label{eq:consistency}
\end{equation}
where, $\phi$ is the increment function of the numerical solution by a given method.

\textit{Definition 2- Stability}: A numerical method is called \textit{stable} if the global error, $\epsilon_i$,  does not grow with the number of steps. The global error is given by equation (\ref{eq:stability}) 
\begin{equation}
\begin{aligned}
\epsilon_i &= x(t_i) - x_i  \leq C.\tau(t_i,x_i,H)
\end{aligned}
\label{eq:stability}
\end{equation}

\textit{Definition 3- Convergence}: A numerical method is called \textit{convergent} if the global error  $\epsilon$ $\rightarrow$ 0  as  $H$ $\rightarrow$ 0.

\textit{Lax Theorem}: If a method is \textit{consistent} and \textit{stable}, it is convergent \cite{Holmes2006IntroNumerical}.
\textit{Consistency} + \textit{Stability} $\Rightarrow$ \textit{Convergence}

\subsection{Test System Definition}

In order to study the numerical stability and the convergence behavior of co-simulation
methods a test model of a coupled system has to be defined. In general, the co-simulation methods are applied on non-linear systems. For the purpose of stability and convergence analysis of numerical time integration methods, a linear test model is used following the Dahlquist’s stability theory,. Since coupling requires a minimum of two sub-systems, we first define a  total system of linear \textit{ode} consisting of two state variables, $X_A$ and $X_B$. 
\begin{equation}
\begin{aligned}
\dot{X_A} &= \lambda_A X_A - K_A x_B\\
\dot{X_B} &= \lambda_B X_B + K_B x_A
\end{aligned}
\label{eq:ABsystem}
\end{equation}
where, $\lambda_A < 0, \lambda_B < 0, K_A > 0$ and $K_B > 0$. 

Examination of the eigen values of this system indicates that this system will always be stable with a true solution with initial value of $X_{A0}$ and $X_{B0}$ is given by 
\begin{equation}
\begin{aligned}
\begin{bmatrix} X_A\\ X_B \end{bmatrix} = e^{\small \begin{bmatrix} \lambda_A & -K_A\\ K_B & \lambda_B\end{bmatrix}.t}. \begin{bmatrix} X_{A0}\\ X_{B0} \end{bmatrix}
\end{aligned}
\label{eq:true}
\end{equation}

Now, let us write this same test system  in a coupled system format using Differential Algebraic Equations (DAE)  with inputs $U_A$ and $U_B$ coming from outputs $Y_B$ and $Y_A$ respectively. The DAE for the $A$ sub-system is given by
\begin{equation}
\begin{aligned}
\dot{X_A} &= \lambda_A X_A + U_A\\
Y_A &= K_B X_A
\end{aligned}
\label{eq:DAE-A}
\end{equation}
and the DAE for the $B$ sub-system is given by
\begin{equation}
\begin{aligned}
\dot{X_B} &= \lambda_B X_B + U_B\\
Y_B &= -K_A X_B
\end{aligned}
\label{eq:DAE-B}
\end{equation}
where $U_A$ = $Y_B$ and $U_B$ = $Y_A$.

\subsection{Co-Simulation Algorithms}
\label{sec:cosim_algo}
\begin{figure}[h]
	\centering
	\includegraphics[trim=0.4in 6.0in 4.1in 0.3in,width=3.2in]{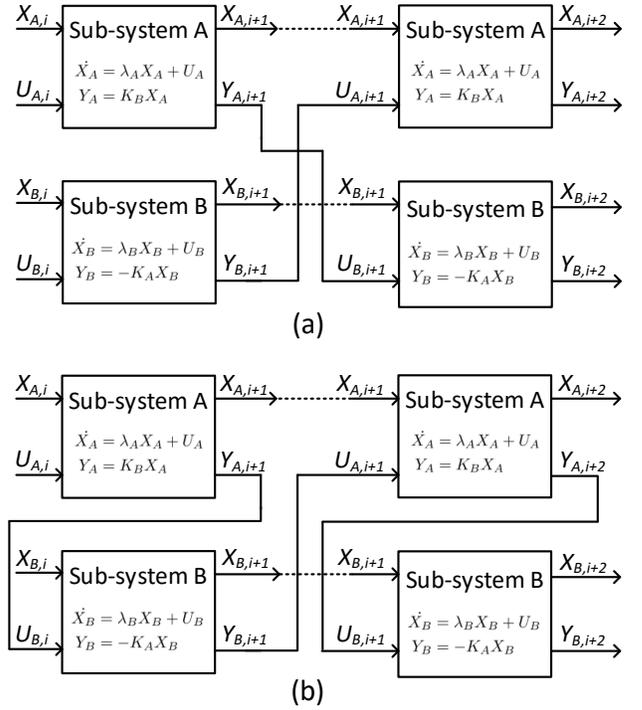} 
	\caption{Co-Simulation block diagrams. (a) Parallel Computation. (b) Series Computation}
	\label{fig:CoupledSimulation}
\end{figure}

The algorithms of the two methods of co-simulation of coupled systems is now discussed in further detail. In both these methods, the key idea is to solve the sub-systems independently and at every integration time step, the input to each of the subsystems is updated from the corresponding output of the other subsystem (Fig. \ref{fig:CoupledSimulation}). The input to the sub-systems during an integration time step is assumed to be constant. 

\subsubsection{\underline{Method 1: Parallel Computation}}~\\
1.~ The initial values of the state variables, $X_{A,i}$, $Y_{A,i}$, $X_{B,i}$, $Y_{B,i}$ are known at $i$ = 0 from the steady state solution. Set the time index $i$ = 0 and the start time, $t_i$ = 0.\\
2.~ Set the input $U_{A,i}$ = $Y_{B,i}$ for the $A$ sub-system.
3.~ Solve the DAEs for the $A$ sub-system.
4.~ Set the input $U_{B,i}$ = $Y_{A,i}$ for the $B$ sub-system.\\
5.~ Solve the DAEs for the $B$ sub-system. \\
6. Increment the time index, $i$, go back to step 2 proceed to the next simulation time step until final time is reached.\\
Since $A$ and $B$ sub-systems use independent solvers, the algorithm lends itself to parallel computation.\\

\subsubsection{\underline{Method 2: Series Computation}}~\\
1.~ Set the time index $i$ = 0 and the start time, $t_i$ = 0. The initial values of the state variables, $X_{A,i}$, $Y_{A,i}$, $X_{B,i}$, $Y_{B,i}$ are known at $i$ = 0 from the steady state solution.\\
2.~ Set the input $U_{A,i}$ = $Y_{B,i}$ for the $A$ sub-system.\\
3.~ Solve the DAEs for the $A$ sub-system.\\
4.~ Set the input $U_{B,i}$ = $Y_{A,i+1}$ for the $B$ sub-system.\\
5.~ Solve the DAEs for the $B$ sub-system.\\
6. Increment the time index, $i$, go back to step 2 proceed to the next simulation time step until final time is reached.\\
Here, in this method, although $A$ and $B$ sub-systems are computed in series, it is still possible to perform parallel computation when multiple $B$ sub-systems exist. 
\subsection{Formulation of Difference Equations}
\label{sec:For_diff}
\subsubsection{Total System Simulation}
The implicit trapezoidal integration method is a very widely used A-stable solver and so this is used to solve the total system to form a baseline for comparison of the coupled numerical methods. The difference equations for the system of equations shown in equation (\ref{eq:ABsystem}) is given by
\small
\begin{equation}
\begin{aligned}
X_{A,i+1} &= X_{A,i} &+ H[0.5(\lambda_A X_{A,i} - K_A X_{B,i})\\ 
~     &~         &+ 0.5(\lambda_A X_{A,i+1} - K_A X_{B,i+1})]\\
X_{B,i+1} &= X_{B,i} &+ H[0.5(\lambda_B X_{B,i} + K_B X_{A,i})\\ 
~	  &~		 &+ 0.5(\lambda_A X_{A,i+1} + K_B X_{A,i+1})]                     
\end{aligned}
\label{eq:diff_T}
\end{equation}
\normalsize
With $\overline{X} = [X_A ~ X_B]'$,  this can now be written of the form
\begin{equation}
\begin{aligned}
\overline X_{i+1} = \overline X_{i}  + H.\phi_T(\overline X_i,H)                  
\end{aligned}
\label{eq:phi_T}
\end{equation}
where $\phi_T(\overline X_i,H)$ is the increment function for the total system using the implicit trapezoidal integration method.\\

\subsubsection{Co-Simulation}
In the coupled system co-simulation, the $A$ sub-system and the $B$ sub-system are different solvers and so could use the same or different numerical methods. The purpose of the analysis is to study the effect of the coupling method considering that the individual solvers are convergent while running independently. Therefore, for the purpose of this study, the implicit trapezoidal method is retained for the $A$ sub-system and an explicit Euler method is used for the $B$ sub-system with a smaller step-size, $h = H/n$, $n$ being the number of small steps.\\

\underline{\textit{Method 1 (Parallel Computation)}}\\
In the parallel computation co-simulation method, from Equations (\ref{eq:true}-\ref{eq:DAE-B}), $U_{A,i}$ = $-K_A X_{B,i}$ and $U_{B,i}$ = $K_B X_{A,i}$. So, we can write the difference equations as
\small
\begin{equation}
\begin{aligned}
X_{A,i+1} &= X_{A,i} &+ H[0.5(\lambda_A X_{A,i} -K_A X_{B,i})\\ 
&    &+ 0.5(\lambda_A X_{A,i+1} - K_A X_{B,i})]\\
X_{B,i+1} &= X_{B,i}(1+h\lambda_B)^n &+ [(\frac{K_B}{\lambda_B}){(1+h \lambda_B)^n -1}]X_{A,i}                    
\end{aligned}
\label{eq:diff_C1}
\end{equation}
\normalsize
This can be expressed of the form
\begin{equation}
\begin{aligned}
\overline X_{i+1} = \overline X_{i}  + H.\phi_{C1}(\overline X_i,H)                  
\end{aligned}
\label{eq:phi_C1}
\end{equation}
where $\phi_{C1}(\overline X_i,H)$ is the increment function for co-simulation method using parallel computation.\\

\underline{\textit{Method 2 (Series Computation)}}\\
In the series computation co-simulation method, as the $A$ sub-system is solved first, $U_{B,i}$ = $K_B X_{A,i+1}$. $U_{A,i}$, however, remains the same as that of Method 1. The difference equations is therefore written as
\small
\begin{equation}
\begin{aligned}
X_{A,i+1} &= X_{A,i} &+ H[0.5(\lambda_A X_{A,i} -K_A X_{B,i})\\ 
&    &+ 0.5(\lambda_A X_{A,i+1} - K_A X_{B,i})]\\
X_{B,i+1} &= X_{B,i}(1+h\lambda_B)^n &+ [(\frac{K_B}{\lambda_B}){(1+h \lambda_B)^n -1}]X_{A,i+1}                    
\end{aligned}
\label{eq:diff_C2}
\end{equation}
This can be expressed of the form
\begin{equation}
\begin{aligned}
\overline X_{i+1} = \overline X_{i}  + H.\phi_{C2}(X_i,H)                  
\end{aligned}
\label{eq:phi_C2}
\end{equation}
\normalsize
where $\phi_{C2}(X_i,H)$ is the increment function for co-simulation method using series computation.\\

\subsection{Convergence Analysis}
As stated earlier in section \ref{sec:Math_Pre}, for a numerical integration method to be convergent, we need to demonstrate consistency and stability. Then, by Lax theorem, the method is convergent. In this section we use the difference equations formulated in the previous section and analyze this criteria to establish the convergence of the co-simulation methods and compare the results with the baseline trapezoidal integration method for the total system.

\subsubsection{Consistency}
For consistency, we are particularly interested in showing that the local truncation error, $\tau_i$, diminishes towards zero as the steps size, $H$ approaches zero. The calculation of truncation from equation (\ref{eq:consistency}), requires the true analytic solution and the numerical increment function.  The analytical solution is given in equation (\ref{eq:true}) and the increment functions of each method are obtained from the difference equations as described in (\ref{eq:phi_T}-\ref{eq:phi_C2}). From these, it can be shown that as $H~\rightarrow$ 0, $\tau_i~\rightarrow$ 0 for the co-simulation methods.

It can also be confirmed graphically by plotting $\tau_i$ for the first time step as $H~\rightarrow$ 0 for two examples of system parameters ($\lambda_A$, $\lambda_B$, $K_A$ and $K_B$). For the base case of implicit trapezoidal integration of the total system the error decay is as expected since this method is known to be consistent. It is also clear that both the co-simulation methods are consistent as well.
\begin{figure}[h]
	\centering
	\includegraphics[trim=0.5in 4.0in 0.5in 0.3in,width=3.3in]{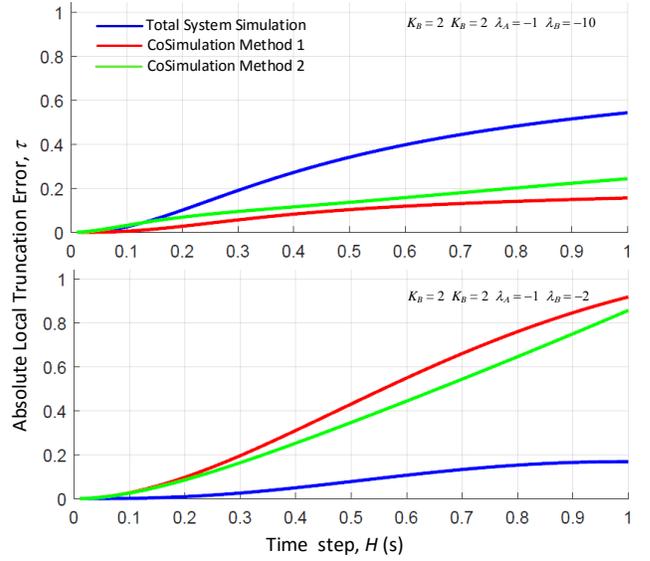} 
	\caption{Comparison of Local Truncation Error at the first time step for Co-simulation methods and Total system simulation.}
	\label{fig:Error}
\end{figure}

\subsubsection{Stability}
For first order linear systems, the stability of the numerical method can be determined when the absolute value of   $\mid m \mid < 1$ when the $x_{i+1}$ is expressed in terms of $x_i$ as equation \ref{eq:m$}.
\begin{equation}
\begin{aligned}
x_{i+1} = m.x_i
\end{aligned}
\label{eq:m$}
\end{equation}
However, for higher order systems, $m$ becomes a matrix, $M$ and so we consider the eigen values of this matrix. If the eigen values are within the unit circle, then the numerical method is stable for the system. 
The stability of the test system can be therefore be analyzed by re-writing the difference equations in Section \ref{sec:For_diff} as 
\begin{equation}
\begin{aligned}
\overline X_{i+1} = M.\overline X_{i}
\end{aligned}
\label{eq:M$}
\end{equation}
and examining the eigen values of $M$. 
For the base case, implicit trapezoidal method, $M$ = 
\small
\begin{equation}
\begin{aligned}
\begin{bmatrix} 1 - 0.5\lambda_A H & 0.5K_A H\\ -0.5 K_B H & 1 - 0.5\lambda_B H\end{bmatrix}^{-1} \begin{bmatrix} 1 + 0.5\lambda_A H & -0.5K_A H\\ 0.5 K_B H & 1 + 0.5\lambda_B H\end{bmatrix}
\end{aligned}
\label{eq:M_IT$}
\end{equation}
\normalsize
For the co-simulation Method 1, $M$ = 
\small
\begin{equation}
\begin{aligned}
\begin{bmatrix} 1 - 0.5\lambda_A H & 0\\ 0 & 1 \end{bmatrix}^{-1} \begin{bmatrix} 1 + 0.5\lambda_A H & -0.5K_A H\\ (\frac{K_B}{\lambda_B})\{(1+h\lambda_B)^n -1\} & (1+h\lambda_B)^n\end{bmatrix}
\end{aligned}
\label{eq:M_C1$}
\end{equation}
\normalsize
For the co-simulation Method 2, $M$ = 
\small
\begin{equation}
\begin{aligned}
\begin{bmatrix} 1 - 0.5\lambda_A H & 0\\ -(\frac{K_B}{\lambda_B})\{(1+h\lambda_B)^n -1\} & 1 \end{bmatrix}^{-1} \begin{bmatrix} 1 + 0.5\lambda_A H & -0.5K_A H\\ 0 & (1+h\lambda_B)^n\end{bmatrix}
\end{aligned}
\label{eq:M_C2$}
\end{equation}
\normalsize

\begin{figure}[h]
	\centering
	\includegraphics[trim=0.5in 3.0in 0.5in 0.3in,width=3.3in]{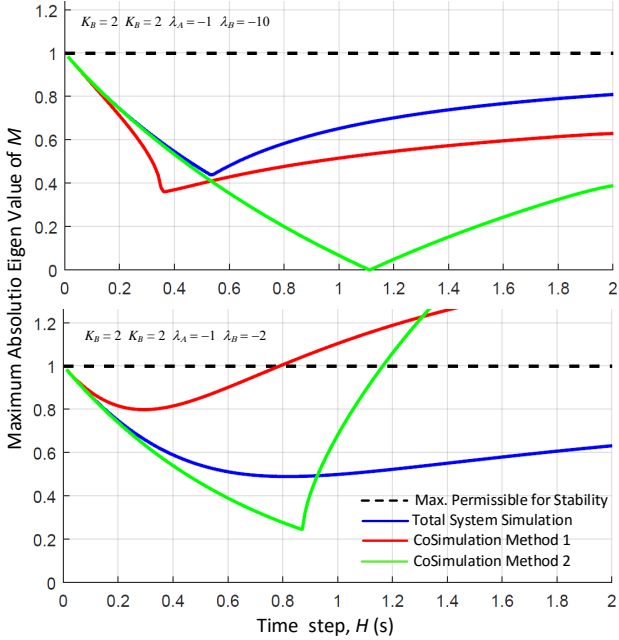} 
	\caption{Maximum absolute eigen value of transformation matrix, $M$. for co-simulation methods and Total system simulation relative to the absolute limit.}
	\label{fig:stability}
\end{figure}

The eigen values of $M$ are not only dependent on the system parameters ($\lambda_A$, $\lambda_B$, $K_A$ and $K_B$), but also on the step size, $H$ and $h$. The base case implicit trapezoidal method is A-stable and so we can expect that the maximum magnitude of the calculated eigen values will be less than 1. However, for the two co-simulation methods, the stability is ascertained for various parameter values and the absolute maximum magnitude of the eigen values for the transformation matrix, $M$ is plotted against $H$. Figures \ref{fig:stability}  show these eigen values for the three simulation methods for the two examples considered.

\subsubsection{Convergence}
By Lax Theorem (Sec \ref{sec:Math_Pre}) convergence follows from consistency and stability. Therefore, the co-simulation methods are convergent as long as the $H$ is chosen to be small enough for the eigen values of the system matrix $M$ to be within the unit circle. This will be further demonstrated by applying the numerical method to compute the discrete evolution of the system state variables in time domain. 

\begin{figure}[t]
	\centering
	\includegraphics[trim=0.5in 4.0in 0.8in 0.3in,width=3.3in]{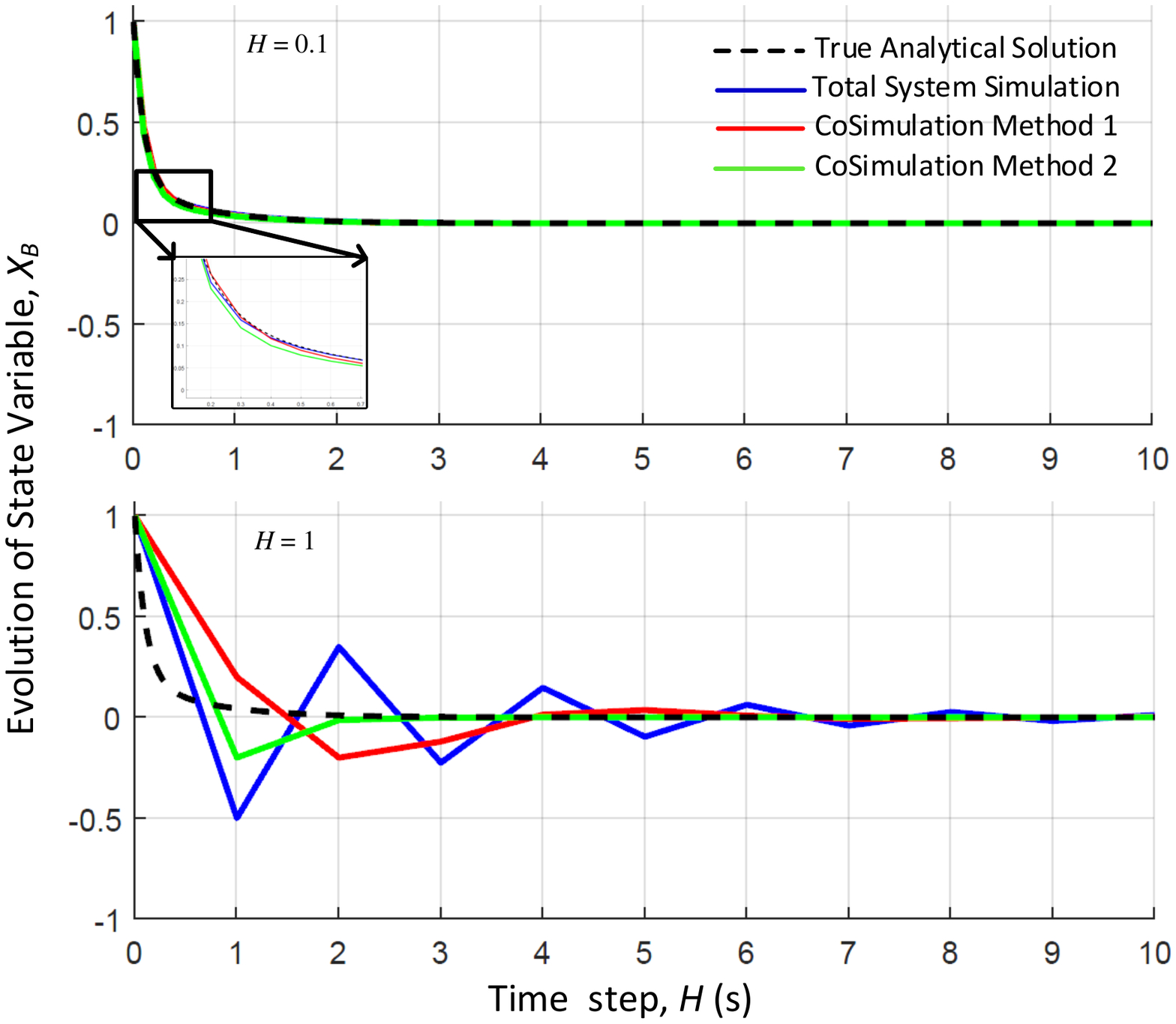} 
	\caption{Discrete evolution of the system using co-imulation and total system simulation compared with the true solution for ($\lambda_A$ = -1, $\lambda_B$ = -10, $K_A$ =2 and $K_B$ =2).}
	\label{fig:Convergence_1}
\end{figure}

Fig. \ref{fig:Convergence_1}(a) shows the results for the first example with $\lambda_A$ = -1, $\lambda_B$ = -10, $K_A$ =2 and $K_B$ =2. When we set the step size, $H$ = 0.1, it can be clearly inferred from the plot that both the the total system solution and the co-simulation methods match very closely with the true solution. However, when the step size is increased to $H$ = 1, the solution takes a longer time to converge. This is evident from the eigen value plots in Fig. \ref{fig:stability}.

\begin{figure}[t]
	\centering
	\includegraphics[trim=0.5in 4.0in 0.5in 0.3in,width=3.3in]{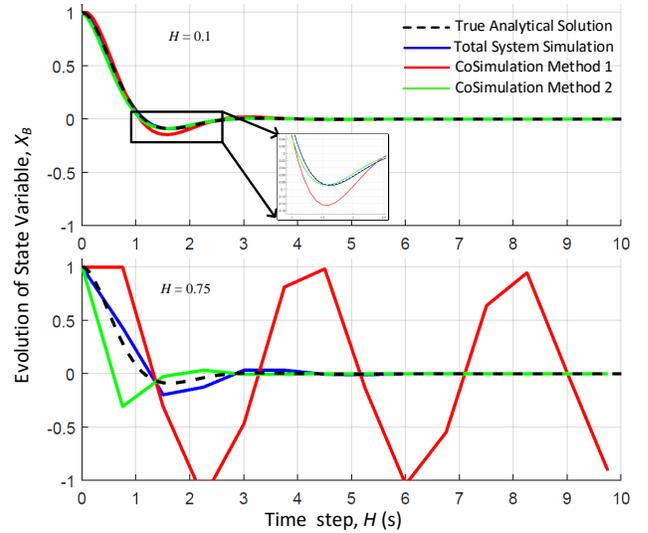} 
	\caption{Discrete evolution of the system using co-simulation and total system simulation compared with the true solution for ($\lambda_A$ = -1, $\lambda_B$ = -2, $K_A$ =2 and $K_B$ =2).}
	\label{fig:Convergence_2}
\end{figure}

Now, let us consider the second example ($\lambda_A$ = -1, $\lambda_B$ = -2, $K_A$ =2 and $K_B$ =2) where the maximum eigen value magnitude crosses the unity limit in the co-simulation methods. We first set the time step, $H$ to 0.1 and then to 0.75. The corresponding discrete time domain evolution plots are shown in Fig.\ref{fig:Convergence_2}. For $H$=0.1, the results are of the all the simulation results are convergent and follow the true solution.  However, with $H$ increased to 0.75, the eigen value of the $M$ for the co-simulation method 1 is 
almost unity whereas that of method 2 is significantly lower than unity. This would suggest that at this time step, the method 1 is getting dangerously close to instability and hence non-convergent. This is validated in Fig.\ref{fig:Convergence_2}, where the method 1 shows wild oscillations whereas the method 2 is highly stable and convergent towards the true solution. The total system solution, as expected, is stable and convergent in both the cases.

Therefore, from this analysis we can observe that co-simulation methods are stable and convergent as long as we keep the step size small enough so that the maximum eigen value magnitude of the transformation matrix is lower than 1. This limitation is due to the methodology of the co-simulation even though the original system when simulated as a single total system of equations is numerically stable and converges to the true analytical solution.  The coupling terms and the eigen values of the original system influence the convergence of the coupled systems. The analysis also indicates that series computation of the coupled systems has better stability characteristics compared to parallel computation.

\section{Combined Transmission-Distribution System (CoTDS)  Dynamic Co-simulation}
\label{sec:CoTDS}

\begin{figure}[h]
	\centering
	\includegraphics[trim=0.35in 9.4in 3.5in 0.4in,width=3.3in]{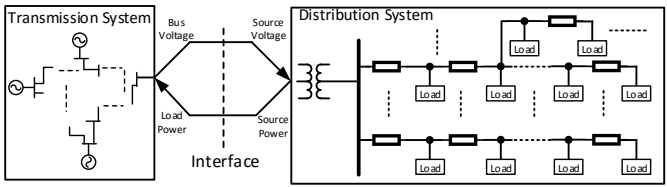}
	\caption{Combined Transmission-Distribution System setup}
	\label{fig:CoTDS1}
\end{figure}

The two methods of co-simulation are now utilized in the dynamic study of transmission and distribution systems connected to each other at the interfacing system bus.
A straightforward way of implementing the CoTDS simulation using either of the co-simulation methods is to represent the transmission system as the $A$ sub-system and the distribution system as the $B$ sub-system with the transmission system bus where the distribution feeder originates as the point of coupling. At this point, the load power of the transmission system is its input and the bus voltage its output. In contrast, the source voltage of the distribution system becomes its input and correspondingly the source power becomes the output. The co-simulation of the CoTDS as shown in Fig. \ref{fig:CoTDS1} is further elaborated in this section.

\subsection{Assumptions and Scope}
For the purpose of the study in this paper, it is assumed that the distribution system load at the sub-station end is balanced. Unbalance in distribution system load is handled through node-level dynamic components and three-phase power flow which will be discussed in the subsequent sections on CoTDS co-simulation. The objective of this study is to demonstrate the co-simulation methods to utilize existing distribution system power flow software in CoTDS co-simulation. This methodology can be further developed to handle unbalanced systems.

\subsection{Transmission and Distribution Dynamic Modeling}

The transmission system dynamic model is comprehensively studied in literature and is well documented in \cite{Khaitan2013high}.  The power system is mathematically represented by a system of differential and algebraic equations (DAEs). The  DAEs in the transmission system dynamic model consist of dynamic components such as generators, exciters, governors and the network. While the network is represented only by algebraic equations, the other components comprise of both differential and algebraic equations. Together they form the DAE for the transmission system. The model is given by the following equation (\ref{eq:fg_trans}) with $x_T$ and $y_T$ as the set of transmission system differential and  algebraic state variables respectively. $x_T$ contains variables related to generator dynamics including the exciter and governor control. And $y_T$ contains the transmission network variables of bus voltages, generator powers and the exciter and governor references. $u_T$ is the set of inputs which is the load power at the load buses where the load is represented by the source power of the distribution system. The corresponding bus voltages at these load buses are the inputs to the distribution system of equations.
\begin{equation}
\begin{aligned}
\dot{x_T} &= f_T(x_T,y_T,u_T)\\
0         &= g_T(x_T,y_T,u_T)
\end{aligned}
\label{eq:fg_trans}
\end{equation}
 
The details of the distribution system dynamic model is presented in \cite{venkatraman2017combined}.  
It has loads comprising of various load components such as static loads (ZIP loads), induction motor loads and reactive shunt compensators. The nodes can also include distribution generator (DG) inverters feeding power into the distribution network supporting a fraction of the total distribution system load.
In this paper, we are not considering the DG inverter model as it is outside the scope of the paper and will be considered in a future publication. The overall structure of the distribution system is also modeled using the
DAE formulation. The comprehensive non-linear model are given in the following equation.
\begin{equation}
\begin{aligned}
\dot{x_D} &= f_D(x_D,y_D,u_D)\\
0         &= g_D(x_D,y_D,u_D)
\end{aligned}
\label{eq:fg_dist}
\end{equation}
Here, $x_D$ and $y_D$ are the vectors of distribution system differential and algebraic state variables respectively. $u_D$ is the input to the distribution system which is the source voltage at the sub-station entrance of the distribution system. This is the same as the corresponding load bus voltage of the transmission system.
With the exchange input output variables of the two systems thus identified, the two sets of DAEs can now be represented  using the co-simulation methods detailed in the previous section.

\subsection{CoTDS co-simulation algorithm and implementation}
\begin{figure}[t]
	\centering
	\includegraphics[trim=1.0in 6.0in 2.0in 0.25in,width=3.4in]{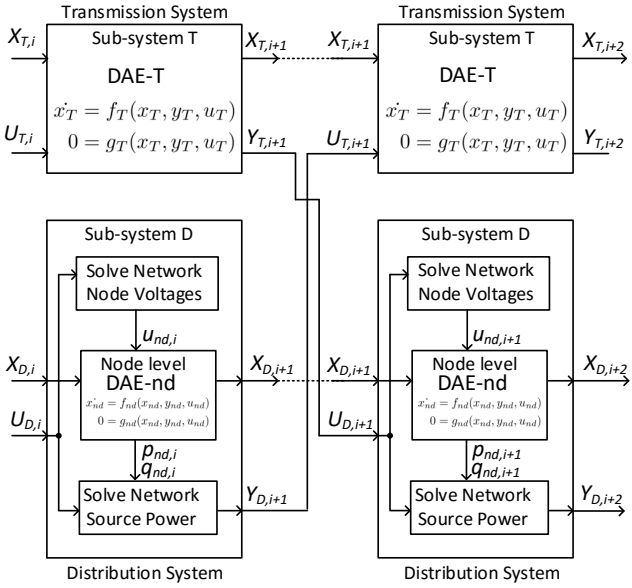}
	\caption{CoTDS co-simulation block diagram using \\ Method 1: Parallel computation}
	\label{fig:CoTDS_Sim_parallel}
\end{figure}

\begin{figure}[t]
	\centering
	\includegraphics[trim=1.0in 6.0in 2.0in 0.25in,width=3.4in]{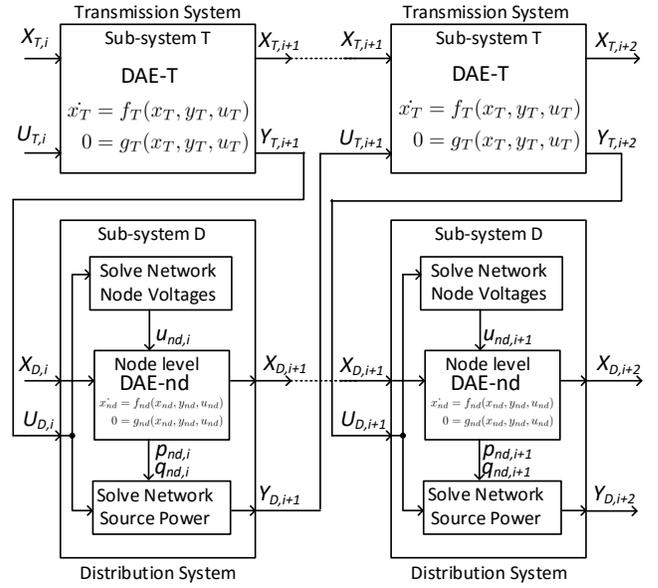}
	\caption{CoTDS co-simulation block diagram using \\ Method 2: Series Computation}
	\label{fig:CoTDS_Sim_series}
\end{figure}

The co-simulation methods as applied to the CoTDS dynamic simulation is proposed in Fig. \ref{fig:CoTDS_Sim_parallel} and Fig. \ref{fig:CoTDS_Sim_series}. 
The transmission system is denoted by Sub-system T and the distribution system is denoted by sub-system D corresponding to sub-systems A and B in the discussions of Sec. \ref{sec:Math}. In the proposed method, the distribution system set of DAEs is solved at a node level instead of solving the complete set of DAEs of an entire distribution system together. The advantages with this approach are:\\
1. A dynamic component can be handled individually using the node voltage as its input and interfaced with the network.\\
2. Existing power flow solvers for the distribution system can be directly used to solve for the network node voltages and source power.

Equation (\ref{eq:fg_dist}) is now written at a node level for each dynamic component as
\begin{equation}
\begin{aligned}
\dot{x_{nd}} &= f_{nd}(x_{nd},y_{nd},u_{nd})\\
0         &= g_{nd}(x_{nd},y_{nd},u_{nd})
\end{aligned}
\label{eq:fg_node}
\end{equation}

The mathematical background of the co-simulation is still applicable as there is no change in the overall scheme of exchanging variable information. Therefore the algorithm presented in Sec. \ref{sec:cosim_algo} is employed to the CoTDS dynamic simulation as follows:

\begin{enumerate}
	\item Solve transmission system  power flow and distribution system  power flow iteratively  \cite{Sun2015master} to set initial values of all variables. The time index, $i$ and the time $t_i$ are initialized to 0.
	\item Set the input $U_{T,i}$ from the source power, $Y_{D,i}$ of the distribution system.
	\item Solve the DAE of the transmission system to obtain $X_{t,I+1}$ and $Y_{t,i+1}$.
	\item For parallel computation, the $U_{D,i}$ is set by $Y_{T,i}$ and for series computation, $U_{D,i}$ is set by $Y_{T,i+1}$.
	\item The distribution system is solved in the following steps.
	\begin{enumerate}
		\item Using the $U_{D,i}$, the power flow is performed on the distribution network  to obtain the node voltages.
		\item The node voltages are passed to the node-level DAE block where the DAE of the dynamic component at each node is solved.
		\item The power at each node is updated and power flow is repeated on the distribution network to obtain the total source power, $Y_{D,i+1}$. 
	\end{enumerate}
	\item Increment $i$ by 1, $t_i$ by the simulation time step and go back to step 2 until final time is reached.
\end{enumerate}

When there are multiple distribution systems, the step 5 of the co-simulation algorithm for all the distribution systems can be applied simultaneously for both the series and the parallel computation methods. Therefore, the benefit of parallel computing of multiple distribution systems can be realized even in case of series computation method of co-simulation.
From the algorithm, it can be noted that the for the transmission system simulation, we can use existing phasor domain software. For the distribution system, we can easily interface a power flow solver by handling the node level component dynamics through an intermediary DAE solver and exchange the input output information through this interface. This is a significant benefit as most distribution system software easily handle power flow and can provide the necessary node voltages and the total source power. So by handling the dynamics of the node-level dynamic components using an intermediate software the combined dynamics of the entire system can very easily be studied without the need for changing the software of either of the simulators. 

\begin{figure}[h]
	\centering
	\includegraphics[trim=0.25in 9.4in 3.4in 0.4in,width=3.0in]{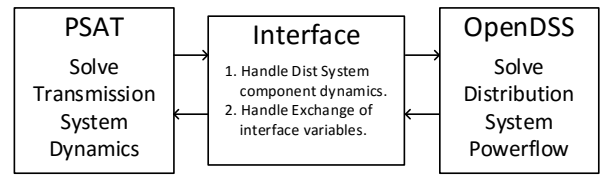}
	\caption{Implementation of CoTDS co-simulation}
	\label{fig:CoTDS2}
\end{figure}

The implementation of the CoTDS co-simulation is demonstrated using PSAT \cite{psat} as the transmission system simulator and  OpenDSS \cite{OpenDSS} as the distribution system power flow solver as shown in Fig. \ref{fig:CoTDS2}. The interface software is implemented in MATLAB. This approach does not require modification of the either PSAT or OpenDSS solvers and therefore this methodology can very easily be extended to other similar platforms.

\section{Validation of CoTDS Dynamic Co-simulation}
\label{sec:validation}
In section, \ref{sec:Math} it has been shown that the co-simulation of two coupled systems are numerically stable and convergent as long as the step size is kept small. If the step size is large, although the actual system is stable, the numerical results can be highly unstable. In this section, we validate this result on the CoTDS co-simulation against the total system simulated in a single dynamic solver. In addition further validation of the CoTDS co-simulation is performed against Simscape EMTP simulation to demonstrate the effectiveness of the proposed CoTDS co-simulation approach.

\subsection{Validation of co-simulation against PSAT simulation for total system}
\label{sec:validationPSAT}
In this section, a test case is setup to simulate a dynamic event first using PSAT which uses implicit trapezoidal integration to solve the total system dynamic equations and provide a reference behaviour for validating the co-simulation methods. The co-simulation is setup using methods 1 and 2 as described in the previous section. The node level component dynamics in the distribution systems are performed using $ode45$ which is an explicit method readily available in MATLAB. 

\begin{figure}[h]
	\centering
	\includegraphics[trim=0.75in 6.2in 3.4in 0.4in,width=3.2in]{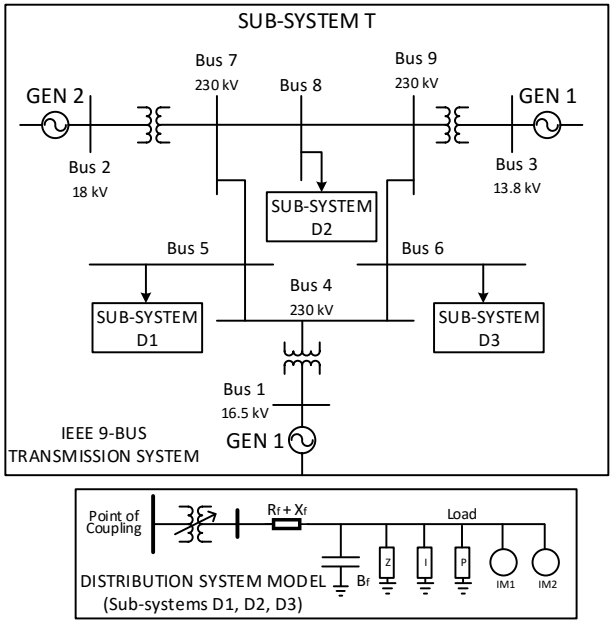}
	\caption{Test-case 1 for validation of the CoTDS co-simulation}
	\label{fig:TestSystem1}
\end{figure}

The test system for studying the proposed co-simulation approach is shown in Fig. \ref{fig:TestSystem1}. 
A WECC 9-bus transmission system is interconnected with aggregated distribution systems at the load buses (5, 6 and 8). The distribution system loads are represented by a combination of static loads, induction motor loads and a lumped distribution feeder impedance 
 
In this test case, since there are three load buses, we thereby have sub-system D1, D2 and D3 for the distribution system. In each of the sub-systems D1, D2 and D3, the loads are comprised of 70\% static load and 30\% induction motor loads. The static loads are further divided into constant impedance ($Z$), constant current ($I$) and constant power ($P$). The induction motor loads are split into two motors, IM1 (60\% fraction) and IM2 (40\% fraction).

A dynamic event is created by initially setting the IM2 on Bus 6 as disconnected from the network and connecting it at $t$ = 11s. The start up of the induction motor creates a transient dip in the bus voltage due to the motors drawing a large amount of reactive power. This event is analyzed using the proposed co-simulation methods and comparing it against total system simulation with PSAT.

\begin{figure}[h]
	\centering
	\includegraphics[trim=0.5in 0.3in 2.25in 0.5in,width=3.1in]{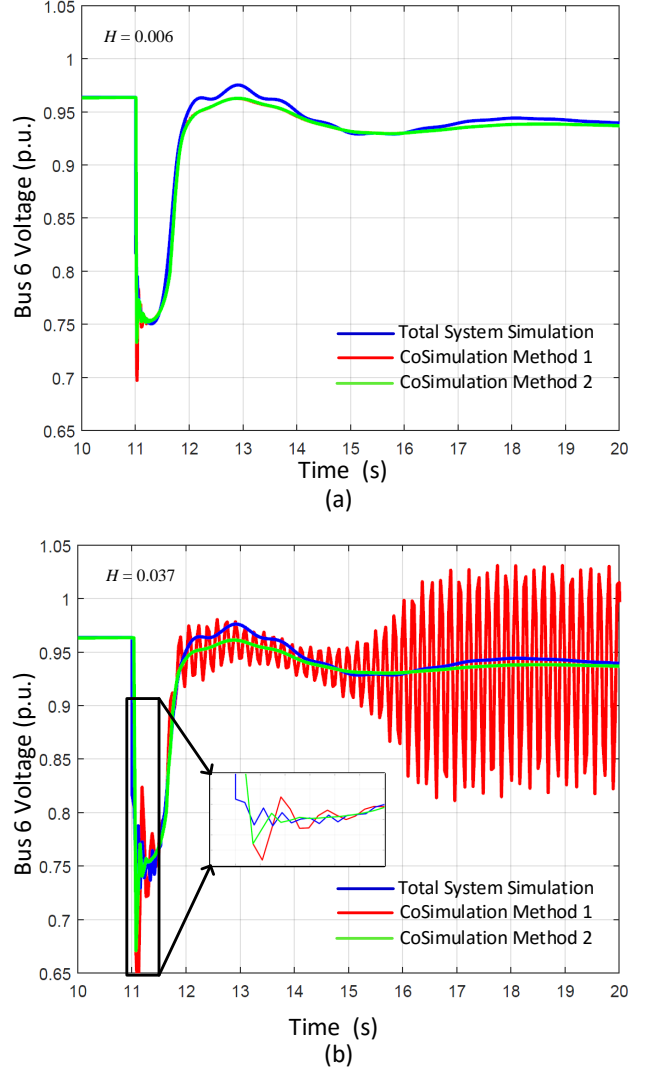} 
	\caption{Bus Voltage evolution during induction motor startup for (a) $H$ = 0.006s and (b) $H$ = 0.037s.}
	\label{fig:Validation1_result}
\end{figure}

Fig. \ref{fig:Validation1_result}(a) shows the simulation result with a small time step of $H$ = 0.006s. This shows both co-simulation methods to give almost identical results and the voltage evolution matches the result obtained from simulating the entire system in PSAT. However, when a higher time step is used, $H$ = 0.037s, Fig. \ref{fig:Validation1_result}(b) the co-simulation method 1 displays numerical stability issues. The co-simulation method 2 shows a stable and convergent result even at a higher time-step. 
This result corresponds to that obtained by rigorous mathematical analysis of the coupled system co-simulation methods in Sec. \ref{sec:Math}. A summary of comparison of the simulation methods in this study is shown in Table \ref{tab:CompareMethods}.

\begin{table}[h]
	\caption{Impact of integration time-step, $H$ on convergence}
	\begin{tabular}{p{6em}  p{6em}  p{6em} p{6em}} 
		\hline\T
		\textbf{Characteristic} & \textbf{Total System Simulation} & \textbf{CoSimulation Method 1}  & \textbf{CoSimulation Method 2} \T\B \\
		\hline \T
		Computation Algorithm & Full-DAE solved together & Parallel ~~~~computation of T and D.& Series ~~~~~~~computation of T and D. \T   \\
		Time-step, $H$ for Stability & Large & Small & Large  \T\B   \\ \T\B
		Convergence at large $H$ & Slow & Diverges & Fast  \\
		\hline
	\end{tabular}
	\label{tab:CompareMethods}
\end{table}

\subsection{Validation of co-simulation against Simscape EMTP simulation for total system}
\label{sec:validationEMTP}
In this section, the  proposed CoTDS co-simulation using the method 2 is further validated against commercially available Simscape Power Systems software. The purpose of this validation is to prove the effectiveness of the co-simulation method by taking a three-phase distribution system and monitoring the behaviour of additional system details like active, reactive power dynamics which cannot be obtained using PSAT. Since Simscape is a complete EMTP three-phase sinusoidal simulation it provides an accurate performance reference for the validation.

\begin{figure}[h]
	\centering
	\includegraphics[trim=0.25in 8.9in 4.4in 0.25in,width=3.4in]{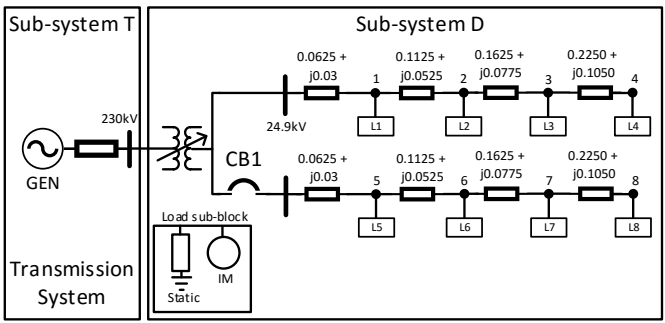}
	\caption{Test case 2 for validation of co-simulation against Simscape. All lines are 3-phase lines, but represented as single line.}
	\label{fig:validationtest2}
\end{figure}

The setup shown in Fig. \ref{fig:validationtest2} comprises of an equivalent generator in the transmission system including the automatic voltage regulating exciter dynamics, governor dynamics and the transmission line connecting to the load. 
The load is represented by a distribution system with two feeders each with 4 nodes. Each node contains a combination of 75\% static and 25\% induction motor loads. The nominal load at each node is evenly distributed amongst the nodes within the feeder.

The simulation is set up as follows: Initially one feeder representing 60\% of the total load of 100MW, 33MVAR in the distribution system is connected to the load bus. The other feeder representing the remaining 40\% of the load is connected at time $t$ = 1s. The transient behaviour of the power up of the feeder is observed using EMTP method and the proposed  co-simulation method 2.

\begin{figure}[h]
	\centering
	\includegraphics[trim=0.25in 4.3in 0.25in 0.25in,width=3.4in]{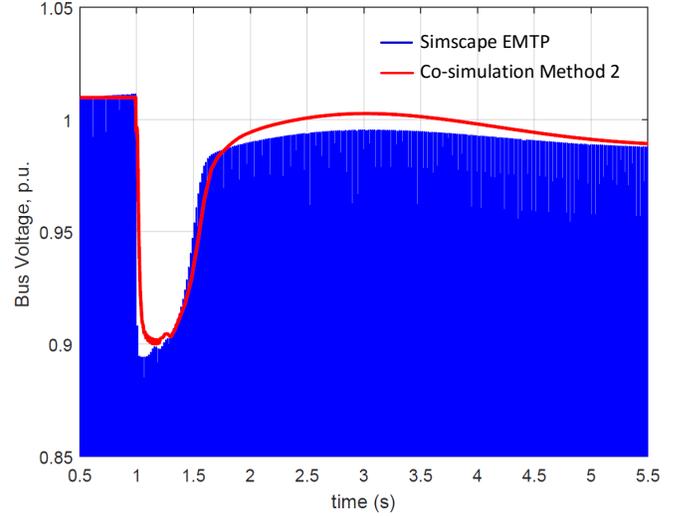}
	\caption{Interface Bus Voltage dynamic behaviour.}
	\label{fig:validationresult1}
\end{figure}

\begin{figure}[h]
	\centering
	\includegraphics[trim=0.25in 1.35in 0.25in 0.25in,width=3.4in]{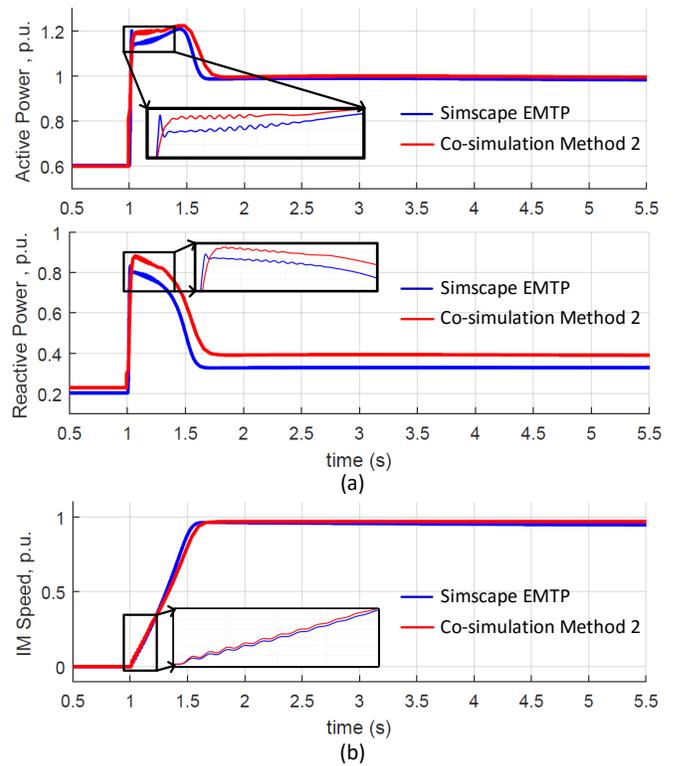}
	\caption{(a) Interface Bus Active and Reactive Power. (b) Induction Motor Speed at Node 1.}
	\label{fig:validationresult2}
\end{figure}

Fig. \ref{fig:validationresult1} shows load bus voltage transient behavior during the connection of the feeder to the system. The EMTP simulation shows the complete transient in full detail with the actual instantaneous voltage plotted relative to the system base peak voltage. As the feeder is connected, the  bus voltage at the interface bus dips and recovers due to the heavy reactive power demanded by the induction motor at start up. 

For both the methods, the active and the reactive power variation during the transient event is plotted in Fig. \ref{fig:validationresult2}(a) and speed of the induction motor load is shown in Fig. \ref{fig:validationresult2}(b). 
The CoTDS simulation results using the proposed co-simulation method 2 displays excellent co-relation with the reference EMTP results. The voltage dip magnitude as well as the active and reactive power variation during the feeder connection shows similar behaviour.

\section{Conclusions}
\label{sec:Conclusion}
In this paper, a rigorous mathematical analysis on convergence of numerical methods in co-simulation is presented. Both the series computation and parallel computation methods of co-simulation are shown to be stable and convergent for smaller integration step sizes and they closely match the true analytic solution. For larger step sizes, even if the individual sub-systems are convergent, the co-simulation may not be convergent. The actual step size for convergence has a dependency on the coupling and the system eigen values. The series computation method permits the use of a larger step size relative to that of parallel computation.

Two methods for co-simulation of CoTDS are proposed using parallel and series computation of the transmission system and distribution systems. The numerical performance of CoTDS co-simulation methods are validated against total system simulation in a single time-domain simulation environment.
The results show correspondence with the theoretical convergence analysis of  the co-simulation methods. Series computation method of transmission and distribution system dynamic models is shown to be numerically stable at larger time steps when the parallel computation method requires smaller time-steps to be stable. At larger time steps, the parallel computation method diverges whereas the series computation method converges. An important benefit of the series computation method is that it converges faster than the total system simulation method.

The parallel computation algorithm although requires smaller time step, it is favorable to parallel computing of transmission and distribution system.  In the series computation algorithm, the computed bus voltages at the various interfacing buses can be used at the same time as the source voltage to the distribution systems and therefore renders itself for parallel computing of all the distribution systems. 

The series co-simulation of CoTDS is further validated against commercial EMTP software and the results show remarkable correspondence.

Another significant advantage of the proposed co-simulation approach for CoTDS dynamic simulation is that 
existing software for transmission dynamics and a power flow solver for three-phase distribution system can be used. The distribution system dynamics are handled using an intermediate software by solving the dynamic equations of the node-level dynamic components and exchanging interface variables at every simulation time-step.

\section{Acknowledgment}
The authors gratefully acknowledge the useful discussions with Dr. Thilo Moshagen at Institute of Scientific Computing, Technische Universit\"{a}t Braunschweig, Germany for the development of the mathematical background during the course of this work.

{\footnotesize
\bibliographystyle{IEEEtran}
\bibliography{references}
}

\end{document}